\documentclass[12pt,preprint]{aastex}
\usepackage{epsfig}
\usepackage{psfig}
\def\msun{M_{\odot}}
\shorttitle{{Dynamical Formation of Millisecond Pulsars in Globular Clusters}}
\shortauthors{Hui et al.}
\begin{document}

\title{Dynamical Formation of Millisecond Pulsars in Globular Clusters}

\author{
C. Y. Hui\altaffilmark{1,2}, K. S. Cheng\altaffilmark{2}, and Ronald
E. Taam\altaffilmark{3, 4, 5} }

\altaffiltext{1} {Department of Astronomy and Space Science, Chungnam National University, 
Daejeon, South Korea}
\altaffiltext{2}
{Department of Physics, University of Hong Kong, Pokfulam Road, Hong
Kong} \altaffiltext{3} {Department of Physics and Astronomy,
Northwestern University, 2131 Tech Drive, Evanston, IL 60208}
\altaffiltext{4} {Academia Sinica Institute of Astronomy and
Astrophysics - TIARA, P.O. Box 23-141, Taipei, 10617 Taiwan}
\altaffiltext{5} {Academia Sinica Institute of Astronomy and
Astrophysics/National Tsing Hua University - TIARA, Hsinchu, Taiwan}

\begin{abstract}
The cumulative luminosity distribution functions (CLFs) of radio millisecond pulsars (MSPs) in globular 
clusters (GCs) and in the Galactic field at a frequency of 1.4 GHz have been examined. Assuming a 
functional form, $N \propto L^q$ where $N$ is the number of MSPs and $L$ is the luminosity at 1.4 GHz, 
it is found that the CLFs significantly differ with a steeper slope, $q=-0.83 \pm 0.05$, in GCs than in 
the Galactic field ($q=-0.48 \pm 0.04$), suggesting a different formation or evolutionary history 
of MSPs in these two regions of the Galaxy. To probe the production mechanism of MSPs in clusters, a search 
of the possible relationships between the MSP population and cluster properties was carried out. The 
results of an investigation of 9 GCs indicate positive correlations between the MSP population and the 
stellar encounter rate and metallicity.  This provides additional evidence suggesting that stellar dynamical 
interactions are important in the formation of the MSP population in GCs.
\end{abstract}


\section{INTRODUCTION \label{intro}}

It is well known that the formation rate per unit mass of low-mass X-ray binaries (LMXBs) is orders 
of magnitude greater in globular clusters (GCs) than in the Galactic field (Katz 1975; Clark 1975).  The 
high formation rate of LMXBs in GCs is attributed to the frequent dynamical interactions in the dense 
stellar environment. As an additional formation channel, the binaries in a GC can also be  
a result of a standard evolutionary path identified for MSP formation in the Galactic 
field. This has stimulated many theoretical and observational studies to investigate the relative 
contribution of these two formation processes of compact binaries in the population of GCs (e.g. Fregeau 
2008; Pooley et al. 2003; Pooley \& Hut 2006).

With the superior sub-arcsecond spatial resolution of the \emph{Chandra X-Ray Observatory}, remarkable 
progress has been made in the understanding of the formation processes of close binaries in GCs. For 
example, Pooley et al. (2003) found a positive correlation between the number of close X-ray binaries 
in GCs and the stellar encounter rate, $\Gamma_{\rm c}$.  Specifically, Pooley et al. (2003) 
found an approximately linear relationship between the number of LMXBs and $\Gamma_{\rm c}$, indicating a 
dependence on the properties of GCs. A similar relationship has also been reported by Gendre et al. (2003)
and taken together with the results of Pooley et al. (2003) provide evidence for the dynamical origin of 
LMXBs in GCs.

Since millisecond pulsars (MSPs) have long been proposed as the descendants of LMXBs, they are also 
expected to have a dynamical origin in GCs. 
Due to the existence of extensive pulsar surveys, 140 MSPs 
have been detected in 26 different clusters and a statistical study of their relationship to cluster
parameters is desirable. 
\footnote{For updated statistics, please refer to http://www2.naic.edu/$\sim$pfreire/GCpsr.html}
However, previous studies were not successful in finding evidence for the dynamical origin of MSPs in 
the clusters due to the lack of a relation between the pulsar population and $\Gamma_{\rm c}$ 
in the GCs (e.g. Ransom 2008). 
This can be ascribed to the observational bias in the pulsar searches.  As the distance 
of the GCs spans a rather wide range (cf. Harris 1996), the sensitivity of the observations can differ 
in the searches toward different clusters and hence induce selection effects in the observed sample 
(Ransom 2008). Therefore, the observed number of MSPs is not representative of an unbiased sample for the 
analysis. 

In this paper, we present a method to alleviate the aforementioned problem and investigate the 
possible relationship between the number of MSPs and the cluster properties. In \S2, an investigation of 
the cumulative luminosity distribution functions of MSPs in a number of selected GCs is carried out. We 
subsequently use the obtained results in a correlation analysis in \S3 and discuss the physical 
implications of the possible correlation in \S4.

\section{CUMULATIVE RADIO LUMINOSITY FUNCTIONS \label{clf}}

Among the GCs, nine contain more than three MSPs for which their radio flux densities are reported. This 
enables one to create the cumulative luminosity distribution functions (CLFs) of MSPs for these clusters 
and to estimate their logarithmic slopes by a power law fit. The physical properties of these nine selected 
GCs are summarized in Table~\ref{gc_info}. As the observations for an individual cluster were conducted at 
different frequencies, we convert all flux densities to the values at 1.4 GHz by assuming a spectral index 
of $-1.8$, which is the mean value reported by Maron et al. (2000). 

We model the CLFs with a form of $N(>L_{\rm 1.4~GHz})=N_{0}L_{\rm 1.4~GHz}^q$, where $N_{0}$ represents
the number of MSPs with the pseudo radio luminosity $>$ 1 mJy~kpc$^{2}$ at 1.4~GHz. 
Since no obvious turn-off is observed from the distribution of individual cluster, we have 
taken the entire sample into account.  Following Hessels et al. (2007), we adopt the 
square root of $N(>L_{\rm 1.4~GHz})$ as the uncertainties in the analysis and  
fit $\log N(>L_{\rm 1.4~GHz})$ versus 
$\log L_{\rm 1.4~GHz}$ with a linear regression analysis that take the uncertainties into account. 
The best-fit parameters are tabulated in Table~\ref{gc_lf_par}. 
It can be seen that the best-fit values of $q$ lie in a range $\sim -0.6$ to $\sim -1.6$ with the steepest slope 
inferred for the MSPs in M~3. In this analysis, we have also taken the 
uncertainties associated with the cluster distance determination into account.  In reviewing different 
methods in determining the distances to GCs, it has been concluded that these various methods imply an 
uncertainty of $\pm6\%$ in the cluster distance (Krauss \& Chaboyer 2003; Chaboyer 2008). Adopting the 
uncertainty of distance in constructing the CLFs, this results in an additional error in the normalization 
beyond the statistical error from the fits.\footnote{We adopted $68\%$ confidence intervals in all the 
reported analysis.} Both errors are combined in quadrature which are quoted in 
Table~\ref{gc_lf_par} and adopted in subsequent analysis in \S3. 

By using the best-fit CLFs for extrapolation, we estimate the 
number of MSPs in each cluster with $L_{\rm 1.4 GHz}>0.5$~mJy~kpc$^{2}$. 
Almost all the MSPs in GCs considered in this study have their 
radio luminosities above this threshold level. Since the errors of both $q$ and $N_{0}$ 
are considered in the extrapolation, the uncertainties of these estimates are larger than that of $N_{0}$. 
These population estimates are also given in Table~\ref{gc_lf_par}. 
For those clusters which have been searched deep enough that the sensitivity is close to this threshold, 
e.g., 47~Tuc, the population estimates based on the CLFs are close to the observed number of uncovered MSPs. 
On the other hand, for those clusters where pulsation searches have not yet reached this sensitivity level, our 
results provide predictions for their MSP populations when searches towards these clusters become sufficiently deep. 

\begin{figure}
\begin{center}
\psfig{figure=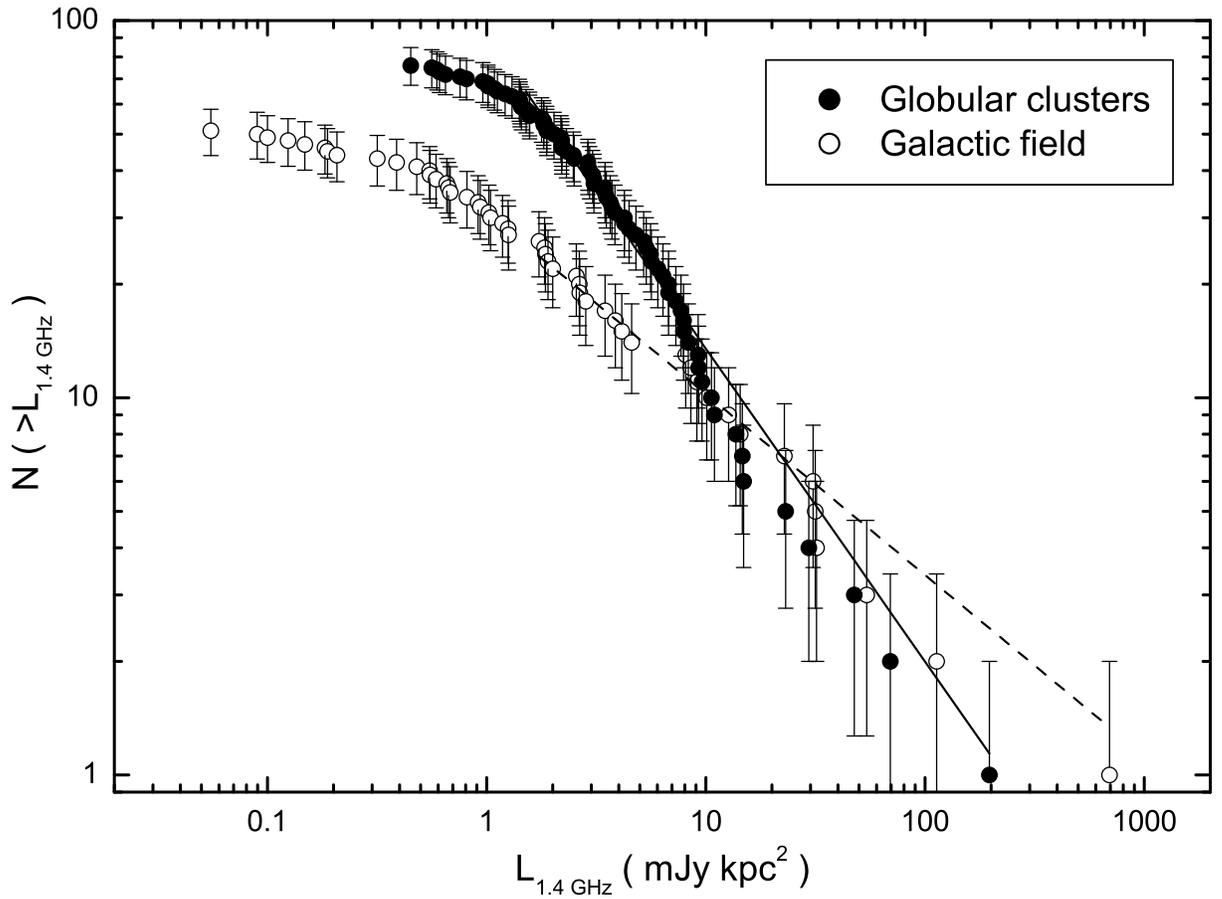,width=18cm,clip=}
\end{center}
\vspace{-1cm}
\caption[]{Comparison of the cumulative radio luminosity distributions of the MSP 
populations in globular clusters and in the Galactic field. The best-fit functions of these two 
populations are shown as the straight lines in the plot.} 
\label{gal_gc_clf}
\end{figure}

Apart from estimating the CLFs for these individual GCs, we have also constructed the luminosity function 
by combining all the cluster MSPs (i.e. 76 pulsars in total) used in this study. The combined CLFs of the 
selected cluster MSP population are displayed in Figure~\ref{gal_gc_clf}. 
In the combined distribution, we have observed there is a turn-off for the pseudo-luminosities smaller 
than $\sim1.5$~mJy~kpc$^{2}$. Hessels et al. (2007) have also observed the same behaviour in an independent 
analysis. In order to compare our result with Hessels et al. (2007),  we followed their procedure and 
used a minimum luminosity cut-off of $L_{\rm 1.4~GHz}=1.5$~mJy~kpc$^{2}$ for the fitting procedure. The results 
of this analysis are summarized in Table~\ref{gal_gc_clf_par}. The logarithmic slope inferred 
from this collective sample is $q=-0.83\pm0.05$, which is consistent with the value deduced by Hessels et 
al. (2007) (i.e. $q=-0.77\pm0.03$) within $1\sigma$ error. 

Hessels et al. (2007) have further compared the luminosity distribution of the MSP population in GCs with those of 
other pulsar populations. They suggested that the distribution of cluster MSPs is marginally consistent with 
that of the MSPs in the Galactic field reported by Corde \& Chernoff (1997) and Lyne et al. (1998). However, the samples
adopted in these investigations contain not more than 22 pulsars. This relative small sample size may introduce 
bias in the analysis and inference. Thanks to the extensive pulsar searches in 
the recent years, the whole pulsar population has been significantly increased. Hessels et al. (2007) have also 
compared their results with a more updated pulsar sample reported by Lorimer et al. (2006) which 
used 1008 pulsars in their 
study. Lorimer et al. (2006) have found a distribution of $d\log N\sim -0.8d\log L$ for their sample, 
which is very similar to that inferred from the cluster population as reported by Hessels et al. (2007) and this 
paper. Nevertheless, the sample used by Lorimer et al. (2006) consist of non-recycled canonical pulsars. 
It is more instructive to compare the CLF of the cluster MSP population and the MSPs in the Galactic field, 
both of which have undergone the recycling processes in binary systems. Therefore, we construct the luminosity 
function for the MSP population in the Galactic field with all the available data in the ATNF catalog 
(Manchester et al.  2005). 

To do this, we specifically selected all pulsars in the Galactic field with $P<$ 
20 ms.  In total, there are 51 MSPs with flux densities reported in the catalog, which enable us to 
construct the CLF.  For comparison, the CLF of the Galactic population is over-plotted along with the 
cluster population in Figure~\ref{gal_gc_clf}.  
To be consistent with the analysis of the cluster population, we used a minimum luminosity cut-off of 
$L_{\rm 1.4~GHz}=1.5$~mJy~kpc$^{2}$ in the fitting. The inferred slope of the Galactic population is 
$q=-0.48\pm0.04$, which is found to be flatter than the value deduced from the cluster population. 

To further investigate the difference between two populations, we separate each population into two 
sub-samples, namely the binary and the solitary MSPs. The results are summarized in Table~\ref{gal_gc_clf_par}. 
\footnote{For PSRs J1342+2822A and J1342+2822C located in M~3, it is still uncertain whether 
they are binary or solitary and therefore they are omitted in the analysis of separate populations.} 
By studying these sub-populations in GCs separately, we deduce the slopes to be $q=-0.73\pm0.08$ and 
$q=-0.89\pm0.11$ for the binary and the solitary MSPs respectively. These values are consistent with those 
inferred by the Hessel et al. (2007) within $1\sigma$ uncertainties.  In the Galactic field, we also 
examine the luminosity functions separately for the binary and solitary MSPs. The slopes inferred 
for the binary and solitary populations are $q=-0.49\pm0.05$ and $q=-0.24\pm0.11$ respectively, 
which are rather different from those inferred from the cluster population. 
For completeness, we have also repeated the above analysis with all the MSPs (i.e., without excluding those 
with their pseudo-luminosities smaller than 1.5~mJy~kpc$^{2}$). The results are also tabulated in 
Table~\ref{gal_gc_clf_par} for the sake of comparison. 

\section{CORRELATION ANALYSIS \label{correlation}}

\begin{figure}
\begin{center}
\psfig{figure=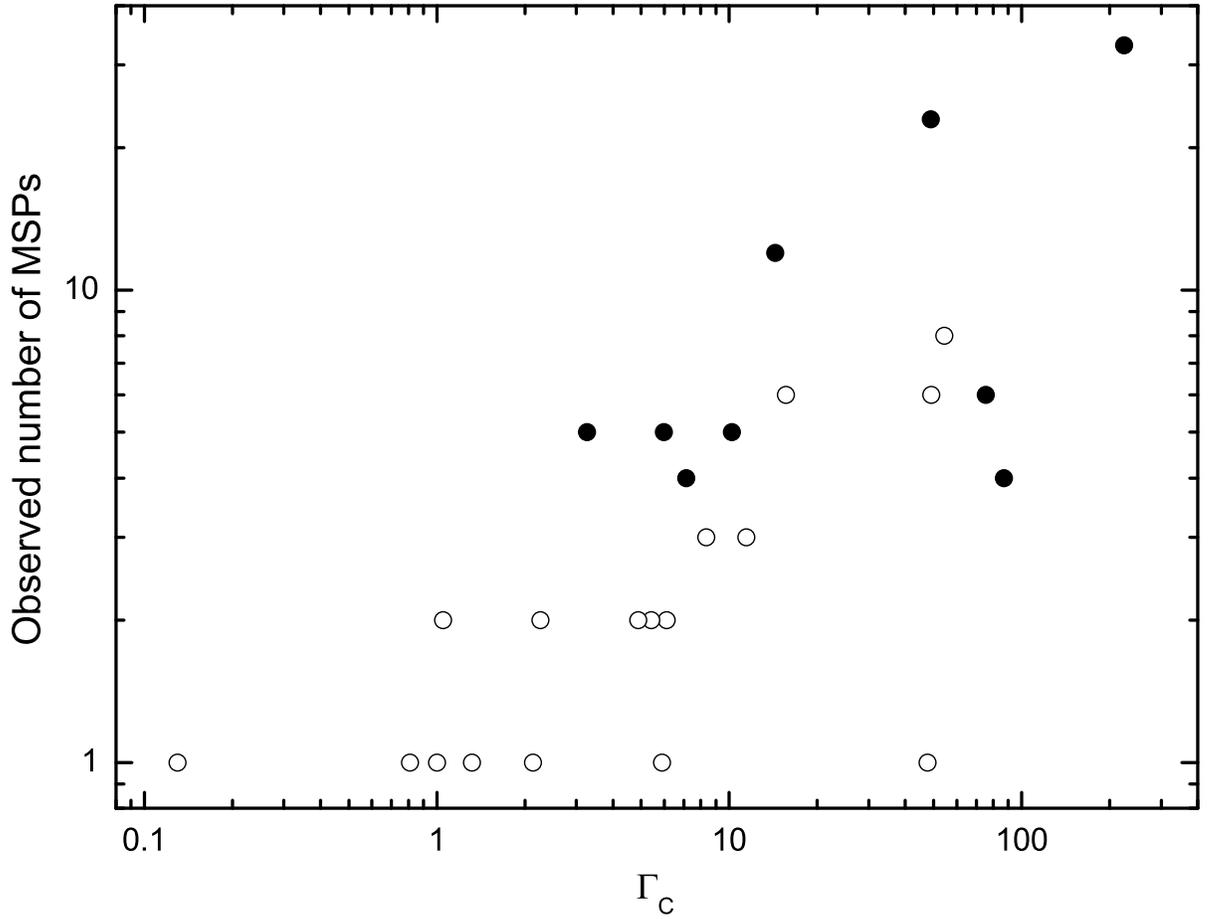,width=18cm,clip=}
\end{center}
\vspace{-1cm}
\caption[]{Observed number of the globular cluster MSPs verse the two-body encounter rate. The filled circles 
represent the nine clusters selected in this investigation and the open circles represent the other MSP-hosting 
clusters.}
\label{obs_msp}
\end{figure}

We have attempted to search for correlations between the observed number of MSPs in each GC with cluster 
parameters. As the most promising correlation is expected with the two-body encounter rate, we begin our analysis 
with this parameter. The two-body encounter rate in a cluster can be estimated as $\Gamma_{\rm c}\propto\rho_{0}^{2} 
r_{\rm c}^{3}\sigma_{0}^{-1}$, where $\rho_{0}$ is the central luminosity density, $r_{\rm c}$ 
is the core radius and $\sigma_{0}$ is the velocity dispersion at the cluster center. In adopting 
the central luminosity density as an estimate of the stellar density at the center, there is an underlying 
assumption that the average luminosities of the stars in the cluster centers are approximately equal 
to $\sim1~L_{\odot}$. Figure~\ref{obs_msp} shows the relation between the observed 
populations with $\Gamma_{\rm c}$. However, no obvious correlations can be identified with this observed sample. 
These two quantities are  correlated at a confidence level of $\sim75\%$ only.  
This conclusion is similar to that reported by Ransom (2008). Owing to the limited amount of telescope time, many 
clusters that host only a single pulsar have not been searched to the same sensitivity level as those of the 
specifically selected targets, such as 47~Tuc. Therefore, the selection effect biases the observed numbers of 
MSPs in different clusters. 

In order to alleviate this problem, we suggest that the use of the CLFs of the investigated clusters. 
With the best-fits of the CLFs (see \S2), we are able to estimate the number of MSP in these 
GCs above a given luminosity threshold and thus obtain an unbiased sample.
Specifically, we take the best-fit values of $N_{0}$ 
in these GCs to estimate the numbers of the MSPs in these clusters with their pseudo-luminosities 
above $>1$~mJy~kpc$^{2}$ and examine whether it is related to different physical quantities of the clusters.  
In this analysis, the possible 
correlation between $N_{0}$ with two-body encounter rate $\Gamma_{\rm c}$, metallicity [Fe/H], cluster 
mass $M_{\rm GC}$, velocity dispersion $\sigma_{0}$ and escape velocity $v_{\rm escape}$ at the cluster 
center are explored. All these quantities are speculated to have influence on the binary formation and hence 
the MSP population in a cluster. 

While the two-body encounter rate $\Gamma_{\rm c}$ is related to 
the binary population resulting from dynamical interactions, the metallicity [Fe/H] of a cluster 
can have a profound influence on the evolution of LMXBs (see Ivanova 2006 and the discussion in \S4). 
On the other hand, if stellar encounters were not the major channel of the binary formation, one would 
expect the binary population to be correlated with the cluster mass $M_{\rm GC}$. 
Assuming a constant mass-to-light ratio, $M_{\rm GC}$ can be estimated from the absolute visual 
magnitude $M_{V}$: $M_{\rm GC}\propto 10^{-0.4M_{V}}$. 
We have also tested the correlation with $\sigma_{0}$ and 
$v_{\rm escape}$ which may possibly be related to the retention of the neutron stars in a cluster. 

\begin{figure}
\psfig{figure=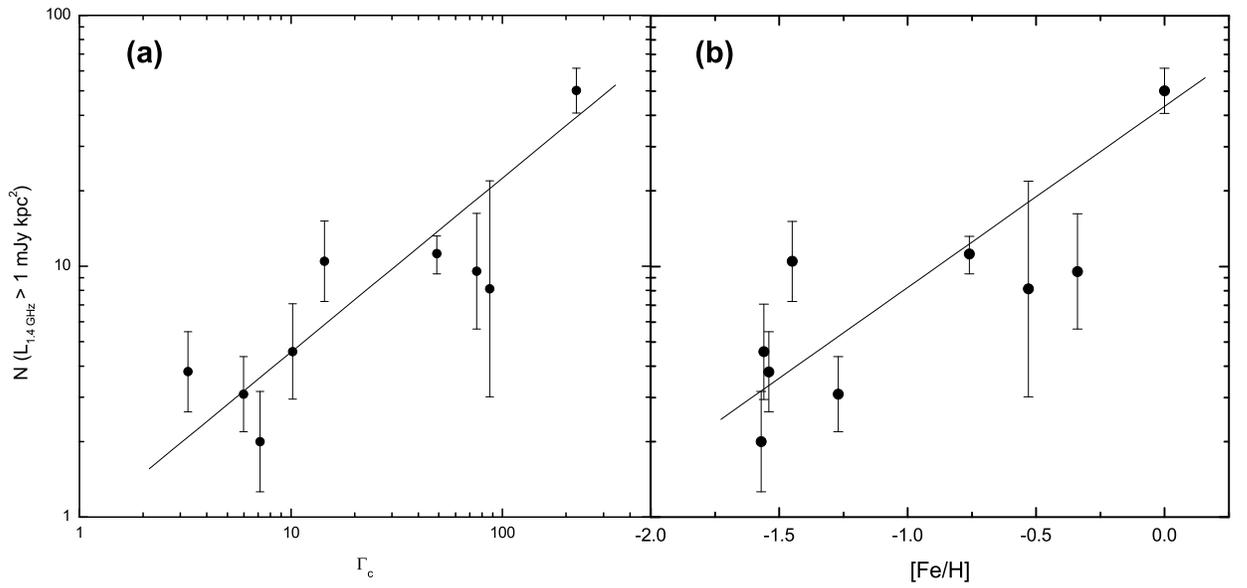,width=18cm,clip=}
\vspace{-1cm}
\caption[]{Number of the globular cluster MSPs with $L_{\rm 1.4 GHz}>1$~mJy~kpc$^{2}$ 
(i.e. $N_{0}$) versus the two-body encounter rate ({\it left panel}) and $N_{0}$ versus the metallicity 
({\it right panel}).}
\label{n_gamma_metal}
\end{figure}

Without a priori knowledge of the distributions of the tested quantities, a nonparametric correlation 
analysis is adopted. The computed Spearman rank correlation coefficients between $N_{0}$ and the various 
quantities are tabulated in Table~\ref{correl}.  Among all the tested quantities, the strongest correlation 
is found between $N_{0}$ and $\Gamma_{\rm c}$. The corresponding Spearman correlation is 0.78 with a chance 
correlation probability of 0.0125. The plot of $N_{0}-\Gamma_{\rm c}$ is displayed in Fig.~\ref{n_gamma_metal}a. 
The correlation between $N_{0}$ and [Fe/H] with a Spearman correlation=0.72 has also been found to be 
significant with a chance correlation probability=0.0298, which is plotted in Figure~\ref{n_gamma_metal}b. 
By taking the errors of $N_{0}$ as the weight in the linear regression analysis, the logarithmic slopes of the 
$N_{0}-\Gamma_{\rm c}$ and $N_{0}-$[Fe/H] relations are found to be $0.69\pm0.11$ and $0.72\pm0.11$ respectively.
For the other tested quantities, there are marginal correlations of $N_{0}$ versus $v_{\rm escape}$ and $\sigma_{0}$ 
at a confidence level $\gtrsim89\%$, though it is not sufficiently significant to secure the relations. 
It is not surprising to note that the rank correlation coefficients are the same for these 
two quantities, as Gnedin et al. (2002) have found that the ratio of $v_{\rm escape}$ to $\sigma_{0}$ has a 
narrow range between $\sim3-5$. 
Among all the tested quantities, the weakest correlation is found for the $N_{0}-M_{\rm GC}$ relation which has a 
chance correlation probability over $60\%$. 

As this choice of luminosity threshold is arbitrary, we further check the robustness of the correlation 
analysis results by repeating the investigation with different thresholds. 
We have repeated the analysis by adopting $N\left(L_{\rm 1.4 GHz}>0.5\right)$ in Table~\ref{gc_lf_par}, 
which provide the estimates 
for the number of MSPs with $L_{\rm 1.4 GHz}>0.5$~mJy~kpc$^{2}$. Almost all the MSPs in GCs considered in this investigation 
have their radio luminosities above this threshold. With these new values, 
the correlations of the MSP number versus $\Gamma_{\rm c}$, [Fe/H], $M_{\rm GC}$, and $v_{\rm escape}$ (or $\sigma_{0}$) are found at 
the confidence levels of $99.47\%$, $92.31\%$, $26.76\%$ and $84.56\%$ respectively. 

For a further test of the robustness, by using the best-fit CLFs in Table~\ref{gc_lf_par}, 
we have also repeated the analysis for a minimum luminosity cut-off 
of $L_{\rm 1.4 GHz}>2$~mJy~kpc$^{2}$. In this case, the correlations with $\Gamma_{\rm c}$, [Fe/H], $M_{\rm GC}$, 
and $v_{\rm escape}$ (or $\sigma_{0}$) are confident at the levels of  $98.32\%$, $98.68\%$, $40.03\%$ and $92.94\%$ 
respectively. Therefore, the degrees of correlation for the tested quantities are found to be insensitive 
to the choice of the threshold. We conclude that the correlation between $\Gamma_{\rm c}$ and the MSP number is 
the most robust among all the tested quanities, which have a confidence level $>98\%$ regardless of the chosen threshold. 
\section{SUMMARY \& DISCUSSION \label{discussion}}

The CLFs of nine GCs, each containing a population of MSPs has been examined. Upon comparison of the MSP   
population in GCs with that in the Galactic field, it has been found that the slopes of the CLFs inferred 
in these two populations significantly differ. 
It is natural to speculate that the CLF is somehow related to the magnetic field and spin of the MSPs. 
Wang, Jiang \& Cheng (2005) have compared the distributions of 
the spin period and the dipolar surface magnetic field for both cluster and disk populations (cf. Fig.~2 and 
Fig.~3 in their paper). Despite the broader distribution for the disk population, their mean values 
are not dissimilar in both populations and therefore cannot solely explain the difference of CLFs. 

Apart from the radio luminosity functions, the X-ray emission properties of the MSPs in the GCs are also found 
to be very different from those in the Galactic field. While the MSPs in the Galactic field generally require 
a hot polar cap component plus a non-thermal power-law tail to model their X-ray spectra (cf. Zavlin 2006), 
the X-rays from a majority of the MSPs in GCs are purely thermal in nature (see Hui et al. 2009 and the 
references therein for a recent review). Cheng \& Taam (2003) suggest the absence of non-thermal X-ray from 
the cluster MSPs can be possibly related to the complicated magnetic field structure. Since the stellar 
interaction in GCs is much more frequent than that in the Galactic field, MSPs in the GCs can possibly  
change their companion several times throughout their lives. Since the orientation of the 
binary after each exchange can differ, the direction of the angular momentum 
accreted during the mass transfer phase subsequent to each exchange can vary possibly affecting the magnetic 
field configuration at the neutron star surface.  Such an evolution could lead to 
a much more complicated magnetic field structure for the MSPs in the GCs than in the case in the Galactic field. 
In such a complicated magnetic field, 
Ruderman \& Cheng (1988) have argued that high energy curvature photons will be emitted and subsequently 
converted into pairs to quench the accelerating region. This provides an explanation for  the absence of non-thermal 
emission in the cluster MSPs. For the same reason, the complicated magnetic field structure can also possibly 
alter the coherent radio emission and result in a different radio luminosity of the cluster MSPs in comparison with 
the disk population. 

Adopting the best-fit normalization inferred from the CLFs of individual cluster as an unbiased estimate 
of the number of MSPs, we have further examined the relationships between the pulsar population and the 
physical properties in GCs. We have found the positive correlations of $N_{0}$ versus $\Gamma_{\rm c}$ as well 
as $N_0$ versus [Fe/H] at a relatively high confidence level. A marginal positive correlation between $N_{0}$ 
and $v_{\rm escape}$ is also suggested. Although a high escape speed implies the presence of a deeper 
gravitational potential well and hence a higher neutron star retention, this correlation is not sufficiently 
significant to warrant such an interpretation.  Hence, we do not discuss this relation any further and focus 
on the physical implications of the $N_{0}-\Gamma_{\rm c}$ and $N_{0}-$[Fe/H] relations. 

Due to the different selection effects in the pulsar search surveys, it is not feasible to directly use the 
detected MSP populations in GCs for a statistical analysis.  Instead, we alleviate the problem by taking 
$N_{0}$ as the estimator for the number of pulsars with pseudo radio luminosites at 1.4~GHz larger than 
1~mJy~kpc$^{2}$. With this consideration, we have found a correlation between $N_{0}$ and $\Gamma_{\rm c}$ at a 
confidence level $>98\%$. We have further found that the strength of this correlation is robust and 
independent of the choice of the luminosity cut-off by repeating the analysis with different 
thresholds.
This provides evidence for the dynamical formation of MSPs in GCs. 
For a competing scenario that the MSPs have a binary origin similar to the Galactic field, 
one should expect the number of MSPs 
to scale with the cluster mass, $M_{GC}$, instead of $\Gamma_{\rm c}$. However, we do not find any convincing 
relationship between $N_{0}$ and $M_{GC}$ (see Table~\ref{correl}). The absence of correlation with $M_{GC}$ 
provides additional support for the dynamical formation scenario. Taken together with the difference in 
the X-ray 
luminosity functions of LMXBs in the field and in globular clusters (see Voss et al. 2009; Kim et al. 2009), it is 
likely that the MSPs have different origins/evolutions in globular clusters relative to the Galactic field. 

We note that the logarithmic slope of the power-law fit in the $N_{0}-\Gamma_{\rm c}$ relationship (i.e. 
$0.69\pm0.11$) is not dissimilar to that of the number of X-ray sources versus $\Gamma_{\rm c}$ ($0.74\pm0.36$ 
Pooley et al. 2003). This dependence on the two-body encounter rate suggests a possible 
relationship between the MSP population and close X-ray binaries in GCs. Apart from the whole X-ray 
binary population, Pooley et al. (2003) and Gendre et al. (2003) have also examined the relationship 
for the individual class of LMXBs which has a logarithmic slope of $0.97\pm0.5$. Although the large 
uncertainty of this slope resulting from the limited sample of LMXBs precludes a definitive conclusion 
concerning the link between LMXBs and MSPs, it is consistent with such an interpretation. 

Theoretical arguments (Verbunt \& Hut 1987) suggest that the number of LMXBs is linearly proportional to 
the stellar encounter rate of the cluster, however direct comparison of their relationship with the current 
two-body encounter rate may be misleading. 
As the MSPs are long lived and are produced by the previous generations of LMXBs, they can have a 
different formation rate from the LMXB population currently observed. This point is important since 
the relaxation time at the cluster core is generally longer than the lifetime of LMXBs (cf. Harris 1996). 
Therefore, the continuous mass segregation at the cluster center can result in a evolution of the stellar 
collision frequency and hence a varying formation rate of compact binary systems. Nevertheless, 
the combination of X-ray and HST observations of Cen A (see Jord\'{a}n et al. 2007) indicate that globular 
clusters with LMXBs are characterized by higher stellar encounter rates than those devoid of LMXBs.  

In addition to the $N_{0}-\Gamma_{\rm c}$ relation, we have also found a positive correlation between $N_{0}$ and 
the metallicity of the GCs. It has been noted that observational evidence suggests that bright LMXBs are 
preferably formed in metal-rich clusters in our Milky Way as well as other galaxies (e.g Bellazzini et al. 
1995; Maccarone et al. 2004; Jord\'an et al. 2004). Ivanova (2006) proposes that the absence of the outer 
convective zone in metal-poor main sequence donor stars in the mass range of $0.85\msun$ - $1.25 \msun$, 
in comparison to their metal rich counterparts can be responsible, since the absence of magnetic 
braking in such stars precludes orbital shrinkage, thereby, significantly reducing the binary parameter 
space for the production of bright LMXBs. For the conventional scenario that LMXBs are the progenitors of 
MSPs, the positive correlation between $N_{0}$ and [Fe/H] is not unexpected since the MSP number should 
scale with that of their progenitors. 

While the stellar encounter rate has been widely accepted as a parameter to indicate which clusters 
are likely to host a large MSP population, our study suggests that the metallicity can also be an 
important parameter. To explore this hypothesis, we suggest that pulsar searches be carried out toward 
metal-rich GCs, such as Liller~1 which has the highest metallicity ([Fe/H]=0.22) among all 150 GCs 
in the Milky Way (cf. Harris 1996). Furthermore, its two-body encounter rate is estimated to be comparable 
with that of 47~Tuc. Therefore, according to these parameters, it is very likely to host a considerable 
number of MSPs. With a dedicated search, this hidden population may be revealed.

\acknowledgments KSC was supported by a GRF grant of Hong Kong Government under HKU700908P. 
RET was supported in part by NSF grant AST-0703950 at Northwestern University and by the Theoretical 
Institute for Advanced Research in Astrophysics (TIARA) operated under the Academia Sinica Institute 
of Astronomy \& Astrophysics in Taipei, Taiwan.

\clearpage

\begin{deluxetable}{lcccccccc}
\tablewidth{0pc}
\tablecaption{Properties of 9 selected globular clusters (cf. Harris 1996: 2003 version; Gnedin et al. 2002).}
\startdata
\hline\hline
Cluster Name & $\log\rho_{0}$\tablenotemark{a} & distance & core radius & [Fe/H]\tablenotemark{b} & $M_V$\tablenotemark{c} & $\sigma_{0}$\tablenotemark{d} & $v_{\rm escape}$\tablenotemark{e} & $\Gamma_{\rm c}$\tablenotemark{f} \\
{}  & $L_{\odot}$pc$^{-3}$ & kpc & pc & & & km/s & km/s & \\\hline\hline
Terzan 5  & 5.06 & 10.3 & 0.54 & 0.00 & -7.87 & 12.7 & 50.5 & 223.8\\
47~Tuc   & 4.81 & 4.5 & 0.52 & -0.76 & -9.42 & 16.4 & 68.8 & 49.0 \\
M~28   & 4.73 & 5.6 & 0.39 & -1.45 & -8.18 & 16.3 & 63.8 & 14.4 \\
NGC~6440   & 5.28 & 8.4 & 0.32 & -0.34 & -8.75 & 21.6 & 85.2 & 75.4\\
NGC~6752   & 4.91 & 4.0 & 0.20 & -1.56 & -7.73 & 7.1 & 32.9 & 10.2 \\
M~5   & 3.91 & 7.5 & 0.92 & -1.27 & -8.81 & 11.8 & 47.7 & 6.0 \\
M~13   & 3.33 & 7.7 & 1.75 & -1.54 & -8.70 & 10.3 & 39.1 & 3.3 \\
M~3   & 3.51 & 10.4 & 1.66 & -1.57 & -8.93 & 9.2 & 37.2 & 7.1 \\
NGC~6441  & 5.25 & 11.7 & 0.37 & -0.53 & -9.64 & 25.2 & 102.0 & 87.1 \\
\enddata
\tablenotetext{a}{Logarithm of central luminosity density}
\tablenotetext{b}{Metallicity}
\tablenotetext{c}{Absolute visual magnitude}
\tablenotetext{d}{$1-$D velocity dispersion at the cluster center}
\tablenotetext{e}{$1-$D escape velocity at the cluster center}
\tablenotetext{f}{Two-body encounter rate estimated by $\rho_{0}^{2}r_{c}^{3}\sigma_{0}^{-1}$ with 
the value scaled with that in M4 which has $\rho_{0}=10^{3.82}$~$L_{\odot}$pc$^{-3}$, 
$r_{c}=0.53$~pc and $\sigma_{0}=8.9$~km/s}. 
\label{gc_info}
\end{deluxetable}

\clearpage
\begin{deluxetable}{lccccc}
\tablewidth{0pc}
\tablecaption{The best-fit parameters of the cumulative radio luminosity 
functions of the MSPs in the selected globular clusters}
\startdata
\hline\hline
Cluster Name & $N_{0}$ & $q$ & $N\left(L_{\rm 1.4 GHz}>0.5\right)$\tablenotemark{a} & $N_{\rm MSP}$\tablenotemark{b} & Reference \\\hline\hline
\\
Terzan~5   & $50.12^{+11.54}_{-9.38}$  & $-0.80\pm0.12$ & $87.10^{+30.39}_{-22.53}$ & 33 & 1 \\
\\
47~Tuc   & $11.22^{+1.96}_{-1.89}$  & $-0.82\pm0.19$ & $19.95^{+6.96}_{-5.50}$ & 23 & 2 \\
\\
M~28   &  $10.47^{+4.66}_{-3.23}$ & $-0.74\pm0.26$ & $17.38^{+12.82}_{-7.38}$ & 12 & 3 \\
\\
NGC~6440   & $9.55^{+6.67}_{-3.93}$  & $-0.59\pm0.27$ & $14.45^{+15.06}_{-7.37}$ & 6 & 4 \\ 
\\
NGC~6752   & $4.57^{+2.51}_{-1.62}$  & $-0.93\pm0.50$ & $8.71^{+10.34}_{-4.73}$ & 5 & 5 \\
\\
M~5   & $3.09^{+1.27}_{-0.90}$  & $-0.58\pm0.31$ & $4.57^{+3.37}_{-1.94}$ & 5 & 6 \\
\\
M~13   & $3.80^{+1.69}_{-1.17}$  & $-0.63\pm0.34$ & $5.89^{+4.83}_{-2.65}$ & 5 & 6 \\
\\
M~3   & $2.00^{+1.17}_{-0.74}$  &  $-1.61\pm1.09$ & $6.09^{+14.33}_{-4.31}$ & 4 & 6 \\
\\
NGC~6441   & $8.13^{+13.75}_{-5.11}$  & $-0.76\pm0.52$ & $13.77^{+39.94}_{-9.88}$ & 4 & 7 \\
\\
\enddata
\vspace{0.5cm}
\tablenotetext{a}{Estimated number of MSPs with their pseudo-luminosities larger than $0.5$~mJy~kpc$^{2}$ by 
using the CLFs.}
\tablenotetext{b}{The observed number of uncovered MSPs.}
\\
References: (1) Ransom et al. 2005; (2) Camilo et al. 2000; (3) B\'egin 2006; (4) Freire et al. 2008;  
(5) D'Amico et al. 2002; (6) Hessels et al. 2007; (7) Freire et al. 2008
\label{gc_lf_par}
\end{deluxetable}

\begin{deluxetable}{lccc}
\tablewidth{0pc}
\tablecaption{Comparison of the cumulative radio luminosity
functions between the MSPs in globular clusters and those in the Galactic field. }
\startdata
\hline\hline
Sample & Sample size\tablenotemark{a} & $N_{0}$\tablenotemark{a} & $q$\tablenotemark{a} \\\hline\hline 
\multicolumn{4}{c}{\bf MSP population in globular clusters}\\
\hline
\\
Total ($L_{\rm min}=1.5$/All) & 58 / 76 & $91.20^{+6.52}_{-6.09}$ / $67.61^{+1.57}_{-1.54}$ & $-0.83\pm0.05$ / $-0.58\pm0.03$ \\
\\
Binary ($L_{\rm min}=1.5$/All) & 32 / 41 & $43.65^{+4.21}_{-3.84}$ / $36.31^{+1.71}_{-1.63}$ & $-0.73\pm0.08$ / $-0.56\pm0.05$ \\
\\
Isolate ($L_{\rm min}=1.5$/All) & 26 / 33 & $46.77^{+6.93}_{-6.04}$ / $32.36^{+2.31}_{-2.16}$ & $-0.89\pm0.11$ / $-0.61\pm0.06$ \\
\\
\hline
\multicolumn{4}{c}{\bf MSP population in Galactic field}\\
\hline
\\
Total ($L_{\rm min}=1.5$/All) & 40 / 51 & $30.90^{+1.46}_{-1.39}$ / $27.54^{+0.64}_{-0.63}$ & $-0.48\pm0.04$ / $-0.32\pm0.02$ \\
\\
Binary ($L_{\rm min}=1.5$/All) & 34 / 39 & $27.54^{+1.30}_{-1.24}$ / $24.55^{+1.16}_{-1.10}$ & $-0.49\pm0.05$ / $-0.36\pm0.03$ \\
\\
Isolate ($L_{\rm min}=1.5$/All) & 6 / 12 & $4.37^{+1.52}_{-1.13}$ / $5.13^{+1.04}_{-0.86}$ & $-0.24\pm0.11$ / $-0.29\pm0.07$ \\
\\
\enddata
\vspace{0.5cm}
\tablenotetext{a}{The entries in these columns depend on whether the analysis has taken: only  MSPs with their pseudo-luminosoties 
greater than 1.5~mJy~kpc$^{2}$ / all MSPs with their flux estimates available. }
\label{gal_gc_clf_par}
\end{deluxetable}

\begin{deluxetable}{lcc}
\tablewidth{0pc}
\tablecaption{Spearman rank correlation coefficient of $N_{0}$ versus various 
physical properties of globular clusters}
\startdata
\hline\hline
Parameters & Spearman rank correlation coefficient & Probability\tablenotemark{a} \\\hline\hline
\multicolumn{3}{c}{ With $N\left(L_{\rm 1.4 GHz}>1~{\rm mJy}~{\rm kpc^{2}}\right)$}\\
\hline
$\Gamma_{\rm c}\propto\rho_{0}^{2}r_{c}^{3}\sigma_{0}^{-1}$ & 0.7833 & 0.9875 \\
$\left[{\rm Fe/H}\right]$      & 0.7167  & 0.9702 \\
$M_{GC}\propto 10^{-0.4M_{V}}$  & -0.2000  & 0.3941 \\
$\sigma_{0}$             & 0.5667  & 0.8884 \\
$v_{\rm escape}$         & 0.5667  & 0.8884 \\
\hline
\multicolumn{3}{c}{ With $N\left(L_{\rm 1.4 GHz}>0.5~{\rm mJy}~{\rm kpc^{2}}\right)$}\\
\hline
$\Gamma_{\rm c}\propto\rho_{0}^{2}r_{c}^{3}\sigma_{0}^{-1}$ & 0.8333 & 0.9947 \\
$\left[{\rm Fe/H}\right]$      & 0.6167  & 0.9231 \\
$M_{GC}\propto 10^{-0.4M_{V}}$  & -0.1333  & 0.2676 \\
$\sigma_{0}$             & 0.5167  & 0.8456 \\
$v_{\rm escape}$         & 0.5167  & 0.8456 \\
\hline
\multicolumn{3}{c}{ With $N\left(L_{\rm 1.4 GHz}>2~{\rm mJy}~{\rm kpc^{2}}\right)$}\\
\hline
$\Gamma_{\rm c}\propto\rho_{0}^{2}r_{c}^{3}\sigma_{0}^{-1}$ & 0.7628 & 0.9832 \\
$\left[{\rm Fe/H}\right]$      & 0.7798  & 0.9868 \\
$M_{GC}\propto 10^{-0.4M_{V}}$  & -0.2034  & 0.4003 \\
$\sigma_{0}$             & 0.6272  & 0.9294 \\
$v_{\rm escape}$         & 0.6272  & 0.9294 \\
\enddata
\tablenotetext{a}{The probability that the correlation coefficient is different from zero.}
\label{correl}
\end{deluxetable}

\clearpage

\clearpage

\clearpage

\end{document}